\documentclass{article}
\usepackage{spconf, graphicx}

\usepackage{booktabs}
\usepackage{wrapfig}
\usepackage{lipsum}
\usepackage{verbatim}
\usepackage{algorithm}
\usepackage{algorithmicx}
\usepackage{algpseudocode}
\usepackage{graphicx}
\usepackage{subcaption}
\usepackage{hyperref}
\usepackage{cite}
\usepackage[cmex10]{amsmath}
\usepackage{amsthm}
\usepackage{epstopdf}
\usepackage{caption}
\usepackage{subcaption}

\usepackage{makecell}
\usepackage{arydshln}
\usepackage{multirow}
\usepackage{multicol}
\usepackage{nicefrac}
\usepackage{cuted}
\usepackage{flushend}
\usepackage{tabularx}
\usepackage{pgfplots}

\def\0{{\mathbf 0}}
\def\1{{\mathbf 1}}

\def\x{{\mathbf x}}
\def\y{{\mathbf y}}

\def\A{{\mathbf A}}
\def\B{{\mathbf B}}
\def\C{{\mathbf C}}
\def\D{{\mathbf D}}

\def\G{{\mathbf G}}

\def\H{{\mathbf H}}
\def\I{{\mathbf I}}

\def\L{{\mathbf L}}

\def\P{{\mathbf P}}

\def\ie{{\textit{i.e.}}}

\def\cE{{\mathcal E}}

\def\cG{{\mathcal G}}

\def\cL{{\mathcal L}}

\def\cV{{\mathcal V}}

\def\bPsi{{\boldsymbol \Psi}}

\def\bTheta{{\boldsymbol \Theta}}

\usepackage{color}

\usepackage{mdwmath}
\usepackage{url}
\usepackage{amssymb}

\def\ie{{\textit{i.e.}}}

\newtheorem{theorem}{\textbf{Theorem}}

\newtheorem{corollary}{\textbf{Corollary}}

\title{Mixed Graph Signal Analysis of Joint Image Denoising / Interpolation}
\name{Niruhan Viswarupan$^\dag$, Gene Cheung$^\dag$\thanks{The work of G. Cheung was supported in part by the Natural Sciences and Engineering Research
Council of Canada (NSERC) RGPIN-2019-06271, RGPAS-2019-00110.}, Fengbo Lan$^\ddag$, Michael S. Brown$^\dag$}
\address{$^\dag$York University, Canada ~~~~~~ $^\ddag$Hong Kong Polytechnic University}
\begin{document}
\ninept
\maketitle
\begin{abstract}
A noise-corrupted image often requires interpolation. 
Given a linear denoiser and a linear interpolator, when should the operations be independently executed in separate steps, and when should they be combined and jointly optimized?
We study joint denoising / interpolation of images from a mixed graph filtering perspective: we model denoising using an undirected graph, and interpolation using a directed graph.
We first prove that, under mild conditions, a linear denoiser is a solution graph filter to a maximum a posteriori (MAP) problem using an undirected graph smoothness prior, while a linear interpolator is a solution to a MAP problem using a directed graph smoothness prior. 
Next, we study two variants of the joint interpolation / denoising problem: a graph-based denoiser followed by an interpolator has an optimal separable solution, while an interpolator followed by a denoiser has an optimal non-separable solution. 
Experiments show that our joint denoising / interpolation method outperformed separate approaches noticeably.

\end{abstract}
\vspace{-0.05in}
\section{Introduction}
\label{sec:intro}
Acquired sensor images are typically noise-corrupted, and a subsequent interpolation task is often required for processing and/or display purposes. 
For example, images captured on a Bayer-patterned grid require demosaicing \cite{malvar2004high, menon2011color}, and a perspective image may need rectification into a different viewpoint \cite{xue2019learning}. 
However, image denoisers and interpolators are often designed and optimized as individual components \cite{zhang2017beyond, xing2021end, kokkinos2019iterative}. This leads to a natural question: should these denoisers and interpolators be independently executed in separate steps, or should they be combined and jointly optimized?

We study the joint image denoising / interpolation problem from a \textit{mixed} graph filtering perspective, leveraging recent progress in the \textit{graph signal processing} (GSP) field \cite{ortega18ieee,cheung18} for image restoration \cite{hu16spl,pang17,liu17,bai19tip,vu21,chen21,yoshida22,gharedaghi23}. 
Our work makes two contributions.  First, we prove that, under mild conditions, a linear denoiser is also an optimal graph filter to a \textit{maximum a poseriori} (MAP) denoising problem using an \textit{undirected} graph smoothness prior\footnote{\cite{chan19} proved a similar theorem for linear denoiser, but our proof based on linear algebra is simpler and more intuitive. See Section\;\ref{subsec:denoiser} for details.} \cite{pang17} (Theorem\;\ref{thm:denoiser}), while a linear interpolator is also an optimal graph filter to a MAP interpolation problem using a \textit{directed} graph smoothness prior \cite{romano17} (Theorem\;\ref{thm:inter}).
These two basic theorems establish one-to-one mappings from conventional linear image filters \cite{milanfar13} to MAP-optimized graph filters for appropriately defined graphs.

Considering both denoising and interpolation simultaneously thus naturally leads to a \textit{mixed} graph model with both directed and undirected edges---a formalism that provides a mathematical framework for joint optimization and explains under which scenarios a joint denoising / interpolation approach would be necessary. 
Our second contribution is to study two variants of the joint problem:
i) an undirected-graph-based denoiser followed by a directed-graph-based interpolator has an optimal \textit{separable} solution (Corollary\;\ref{corol:separable}), and ii) a directed-graph-based interpolator followed by an undirected-graph-based denoiser has an optimal \textit{non-separable} solution (Corollary\;\ref{corol:non-separable}). 
In the latter case, we analytically derive the optimal joint solution that comprises derivative denoising and interpolation operators that are easily computable functions of the input interpolator / denoiser. 
Experiments show that using these computed operators for joint denoising / interpolation of test images can outperform separate approaches noticeably.


\vspace{-0.05in}
\section{Preliminaries}
\label{sec:prelim}
\vspace{-0.05in}
\subsection{GSP Definitions}

\vspace{-0.05in}
We define basic GSP definitions \cite{ortega18ieee}.
A graph $\mathcal{G}=(\mathcal{V}, \mathcal{E})$ consists of a node set $\mathcal{V}$ of size $N$ and an edge set $\mathcal{E}$ specified by $(i,j,w_{i,j})$, where $i,j \in \mathcal{V}$ and $w_{i,j} \in \mathbb{R}$ is a scalar weight of an edge $(i,j) \in \cE$ encoding the similarity / dissimilarity between samples at nodes $i$ and $j$, depending on $w_{i,j}$'s sign.
We define an \textit{adjacency matrix} $\A \in \mathbb{R}^{N \times N}$, where $A_{i,j} = w_{i,j}$ if $(i,j) \in \cE$, and $A_{i,j}=0$ otherwise. 
We consider both \textit{undirected} and \textit{directed} graphs.
An undirected graph $\cG$ means $A_{i,j} = A_{j,i}, \forall i,j \in \cV$, and $\A$ is symmetric. 

For undirected graphs, we define a diagonal \textit{degree matrix} $\mathbf{D} \in \mathbb{R}^{N \times N}$, where $D_{i,i}=\sum_{j} A_{i,j}$. 
Given $\mathbf{A}$ and $\mathbf{D}$, we define a \textit{combinatorial graph Laplacian matrix} as $\mathbf{L} \triangleq \mathbf{D}-\mathbf{A}$.
If there exist self-loops, \ie, $w_{i,i} \neq 0, \exists i$, then the \textit{generalized graph Laplacian matrix} $\L_g \triangleq \D - \A + \text{diag}(\A)$ is typically used instead.


\vspace{-0.05in}
\subsection{Graph Smoothness Priors}

\vspace{-0.05in}
There exist several graph smoothness priors in the GSP literature; each assumes signal $\x$ is smooth w.r.t. the underlying graph $\cG$ but is expressed in slightly different mathematical terms. 
The most common is the \textit{graph Laplacian regularizer} (GLR) \cite{pang17}:
\begin{align}
\x^\top \L \x = \sum_{(i,j)\in\cE} w_{i,j}(x_i-x_j)^2,
\label{eq:GLR}
\end{align}
where $\L$ is the combinatorial graph Laplacian for graph $\cG$. 
If edge weights are non-negative, \ie, $w_{i,j} \geq 0, \;\forall (i,j) \in \cE$, then $\L$ is provably \textit{positive semi-definite} (PSD) and $\x^\top \L \x \geq 0, \forall \x \in \mathbb{R}^N$ \cite{cheung18}. 
GLR can be similarly defined using generalized Laplacian $\L_g$ instead of $\L$.
A small GLR means a connected node-pair $(i,j)$ with large edge weight $w_{i,j}$ should have similar values $x_i$ and $x_j$.

Another common graph smoothness prior is the \textit{graph shift variation} (GSV) \cite{romano17}.
First, define row-stochastic adjacency matrix as $\A_{r} \triangleq \D^{-1} \A$. 
We then write GSV as
\begin{align}
\| \x - \A_r \x\|^2_2 &= \|(\I - \A_r)\x\|^2_2 = \| \L_r \x \|^2_2,
\label{eq:GSV} 
\end{align}
where $\L_r \triangleq \D^{-1} \L$ is the random walk graph Laplacian.
GSV \eqref{eq:GSV} can be interpreted as the $\ell_2$-norm difference between signal $\x$ and its shifted version $\A_r \x$, where the row-stochastic adjacency matrix $\A_r$ is the shift operator. 
GSV can be rewritten as $\x^{\top} \L_r^{\top} \L_r \x$, which was called \textit{left eigenvectors of the random walk graph Laplacian} (LERaG) in \cite{liu17}. 
In contrast to GLR \eqref{eq:GLR}, one important characteristic of GSV \eqref{eq:GSV} is that it is well defined even if the graph $\cG$ is directed. 


\vspace{-0.05in}
\section{Linear Denoisers and Interpolators}
\label{sec:graphInterpolate}
\vspace{-0.05in}
\subsection{Denoiser: Undirected Graph MAP Problem}
\label{subsec:denoiser}

\vspace{-0.05in}
We first establish a theorem to relate a linear denoiser $\bPsi \in \mathbb{R}^{N \times N}$ to a MAP optimization problem using an undirected graph smoothness prior. 
Consider a linear denoising operation written as
\vspace{-0.05in}
\begin{align}
\x = \bPsi \y,
\label{eq:denoiser}
\end{align}
where $\x, \y \in \mathbb{R}^N$ are the denoiser output and input, respectively. 
%
%
Consider next the following standard MAP optimization problem for denoising input $\y$, using GLR \eqref{eq:GLR} \cite{pang17} as the signal prior:
\vspace{-0.05in}
\begin{align}
\min_\x \|\y - \x\|^2_2 + \mu \x^{\top} \L \x,
\label{eq:MAP_denoise}
\end{align}
where $\mu > 0$ is a weight parameter.
Assuming graph Laplacian $\L$ is PSD, \eqref{eq:MAP_denoise} is an unconstrained and convex \textit{quadratic programming} (QP) problem with solution
\begin{align}
\x^* = \left(\I_N + \mu \L \right)^{-1} \y,
\label{eq:MAP_sol_denoise}
\end{align}
where $\I_N$ is an $N \times N$ identity matrix. 
Note that coefficient matrix $\I_N + \mu \L$ is provably \textit{positive definite} (PD) and thus invertible. 

We now connect denoiser $\bPsi$ \eqref{eq:denoiser} and the MAP problem \eqref{eq:MAP_denoise}:
\begin{theorem}
Denoiser $\bPsi$ \eqref{eq:denoiser} is the solution filter for the MAP problem \eqref{eq:MAP_denoise} if $\L = \mu^{-1} (\bPsi^{-1} - \I_N)$, assuming matrix $\bPsi$ is non-expansive, symmetric, and PD. 
\label{thm:denoiser}
\end{theorem}

\begin{proof}
Symmetry means $\bPsi$ has real eigenvalues.
Non-expansiveness and PDness mean $\bPsi$ is invertible with positive eigenvalues $0 < \lambda_k \leq 1, \forall k$.
Thus, $\bPsi^{-1}$ exists and has positive eigenvalues $\{\lambda_k^{-1}\}$. 
Condition $\lambda_k \leq 1, \forall k$ means that $1 \leq \lambda_k^{-1}, \forall k$.
Thus, the eigenvalues of $\bPsi^{-1} - \I_N$ are $\lambda_k^{-1}-1 \geq 0, \forall k$.
This implies $\L = \mu^{-1} (\bPsi^{-1} - \I_N)$ is PSD, and thus quadratic objective \eqref{eq:MAP_denoise} is convex.
Taking derivative w.r.t. $\x$ and setting it to zero, optimization \eqref{eq:MAP_denoise} has \eqref{eq:MAP_sol_denoise} as solution. 
Inserting $\L = \mu^{-1} (\bPsi^{-1} - \I_N)$ into \eqref{eq:MAP_sol_denoise}, we get $\x^* = \bPsi \y$, and thus $\bPsi$ is the resulting solution filter.
\end{proof}

\vspace{-0.05in}
\noindent 
\textbf{Remarks}: 
In the general case, PSD matrix $\L_g = \mu^{-1}(\bPsi^{-1} - \I_N)$ corresponding to non-expansive, symmetric and PD $\bPsi$ is a generalized graph Laplacian to a graph with positive / negative edges and self-loops.
Theorem\;\ref{thm:denoiser} states that, under ``mild" condition, a graph filter---solution to MAP problem \eqref{eq:MAP_denoise} with an \textit{undirected} graph smoothness prior---is equally expressive as a linear denoiser $\bPsi$. 

One benefit of Theorem\;\ref{thm:denoiser} is \textit{interpretability}: 
any linear denoiser $\bPsi$ satisfying the aforementioned requirements can now be interpreted as a \textit{graph filter} corresponding to an \textit{undirected} graph $\cG^u$, specified by $\L = \mu^{-1}(\bPsi^{-1} - \I_N)$, given that $\L$ is symmetric. 
In fact, \textit{bilateral filter} (BF) \cite{tomasi98} has been shown to be a graph filter in \cite{gadde13}, but Theorem\;\ref{thm:denoiser} provides a more general statement.

\vspace{-0.05in}
\subsection{Interpolator: Directed Graph MAP Problem}

\vspace{-0.05in}
We next investigate a linear interpolator $\bTheta \in \mathbb{R}^{N \times M}$ that interpolates $N$ new pixels from $M$ original pixels $\y \in \mathbb{R}^M$:
\begin{align}
\x = \left[ \begin{array}{c}
\I_{M} \\
\bTheta
\end{array} \right] \y,
\label{eq:interpolator}
\end{align}
where $\x = [\y^{\top} ~~ \tilde{\x}^{\top}]^{\top} \in \mathbb{R}^{M+N}$ is the length-$(M+N)$ target signal that retains the original $M$ pixels. 

We define a MAP optimization objective for interpolation, similar to previous \eqref{eq:MAP_denoise} for denoising.
Denote by $\H = [\I_{M} ~~ \0_{M,N}]$ a $M \times (M+N)$ \textit{sampling matrix} that selects $M$ original pixels from signal $\x$, where $\0_{M,N}$ is a $M \times N$ matrix of zeros.  
Denote by $\A$ an \textit{asymmetric} adjacency matrix specifying \textit{directional} edges in a directed graph $\cG^d$ for signal $\x$.
Specifically, $\A$ describes edges only from the $M$ original pixels to $N$ new pixels, \ie, 
\vspace{-0.05in}
\begin{align}
\A = \left[ \begin{array}{cc}
\0_{M,M} & \A_{M,N} \\
\0_{N,M} & \0_{N,N}
\end{array} \right].
\label{eq:A_directed}
\end{align}

We now write a MAP optimization objective using GSV \eqref{eq:GSV} as the signal prior:
\begin{align}
\min_{\x} \|\y - \H \x\|^2_2 + \gamma \|\H (\x - \A \x)\|^2_2, 
\label{eq:MAP_inter}
\end{align}
where $\gamma > 0$ is a weight parameter. 
GSV $\|\H (\x - \A \x) \|^2_2$ states that a smooth graph signal $\x$ should be similar to its shifted version $\A \x$, but we evaluate only the $M$ original pixels in the objective. 
Note that \eqref{eq:MAP_inter} is convex for any definition of $\A$. 

We state formally a theorem to connect interpolator in \eqref{eq:interpolator} and the MAP problem \eqref{eq:MAP_inter}. 
\begin{theorem}
The interpolator $[\I_M; \bTheta]$ \eqref{eq:interpolator} is the solution filter to the MAP problem \eqref{eq:MAP_inter} if $M=N$, $\bTheta$ is invertible, and $\A_{M,N} = \bTheta^{-1}$.
\label{thm:inter}
\end{theorem}

\begin{proof}
we rewrite $\H(\x-\A\x) = \H(\I-\A)\x$.
Given \eqref{eq:MAP_inter} is convex for any $\A$, we take the derivative w.r.t. $\x$ and set it to $0$, resulting in

\vspace{-0.1in}
\begin{small}
\begin{align}
\left(\H^{\top} \H + \gamma (\I-\A)^\top \H^\top \H (\I-\A) \right) \x = \H^{\top} \y .
\label{eq:MAP_inter_linSys}
\end{align}
\end{small}\noindent
Given the definitions of $\H$ and $\A$, we can rewrite \eqref{eq:MAP_inter_linSys} as
\begin{align}
\underbrace{\left[ \begin{array}{cc}
(1 + \gamma) \I_M  & -\gamma \A_{M,N} \\
- \gamma \A_{M,N}^\top & \gamma \A^2_{N,N}
\end{array} \right]}_{\C} \x = \H^{\top} \y,
\label{eq:MAP_inter_linSys2}
\end{align}
where $\A^2_{N,N} = \A_{M,N}^\top \A_{M,N}$.

Using a matrix inversion formula \cite{bernstein2009matrix},
\begin{footnotesize}
\begin{align}
\left[ \begin{array}{cc}
\dot{\A} & \dot{\B} \\
\dot{\C} & \dot{\D} 
\end{array} \right]^{-1}
&= \left[ \begin{array}{cc}
\dot{\P} & -\dot{\P} \dot{\B} \dot{\D}^{-1} \\
-\dot{\D}^{-1} \dot{\C} \dot{\P} & \dot{\D}^{-1} + \dot{\D}^{-1} \dot{\C} \dot{\P} \dot{\B} \dot{\D}^{-1}
\end{array} \right] 
\label{eq:inverse}
\end{align}
\end{footnotesize}\noindent 
where $\dot{\P} = (\dot{\A} - \dot{\B} \dot{\D}^{-1} \dot{\C})^{-1}$, we first compute $\dot{\P}$ for coefficient matrix $\C$ in \eqref{eq:MAP_inter_linSys2} as
\begin{align}
\dot{\P} &= \left( (1+\gamma) \I_M - (-\gamma \A_{M,N}) (\gamma \A_{N,N}^2)^{-1} (-\gamma \A_{M,N}^\top) \right)^{-1} 
\nonumber \\
&= \left( (1+\gamma) \I_M - \gamma \A_{M,N} (\A_{M,N}^\top \A_{M,N})^{-1} \A_{M,N}^\top \right)^{-1}
\nonumber \\
&\stackrel{(a)}{=} \left( (1+\gamma) \I_M - \gamma \A_{M,N} \A_{M,N}^{-1} (\A_{M,N}^\top)^{-1} \A_{M,N}^\top \right)^{-1}
\nonumber \\
&= \left( (1+\gamma) \I_M - \gamma \I_M \right)^{-1} = \I_M ,
\end{align}
where in $(a)$ we apply the assumptions that $N = M$ and $\A_{M,N} = \bTheta^{-1}$ and thus invertible.

We can now write $\x$ from \eqref{eq:MAP_inter_linSys2} as

\vspace{-0.1in}
\begin{small}
\begin{align}
\x &= \C^{-1} \H^\top \y = \left[ \begin{array}{c}
\I_M \\
-(\gamma \A_{N,N}^2)^{-1} (-\gamma \A_{M,N}^\top) \I_M 
\end{array} \right] \y \nonumber \\
&= \left[ \begin{array}{c}
 \I_M \\ \A_{M,N}^{-1} (\A_{M,N}^\top)^{-1} \A_{M,N}^\top
\end{array} \right] \y 
= \left[ \begin{array}{c}
\I_M \\
\A^{-1}_{M,N} 
\end{array} \right] \y .
\label{eq:Q}
\end{align}
\end{small}\noindent
Given $\A_{M,N}^{-1} = \bTheta$ by assumption, we conclude that
\vspace{-0.05in}
\begin{align}
\x = \left[ \begin{array}{c}
\I_M \\
\bTheta \end{array} \right] \y .
\end{align}
Hence, $[\I_M; \bTheta]$ is the solution filter to \eqref{eq:MAP_inter}. 
\end{proof}

\noindent
\textbf{Remarks}: 
Theorem\;\ref{thm:inter} states that, under ``mild" condition, a graph filter---solution to MAP problem \eqref{eq:MAP_inter} with a \textit{directed} graph smoothness prior---is equally expressive as a general linear interpolator $\bTheta$.
The requirements for Theorem\;\ref{thm:inter} mean that the $M$ interpolated pixels are linearly independent.

Intuitively, using \textit{bidirectional} edges $(i,j)$ in an \textit{undirected} graph for denoising makes sense; 
the uncertainty in observed noisy pixels $i$ and $j$ means that their reconstructions depend on each other.
In contrast, using \textit{directional} edge $[i,j]$ in a \textit{directed} graph for interpolation is reasonable; original pixel $i$ should influence interpolated pixel $j$ but not vice versa.


\vspace{-0.05in}
\section{Joint Denoising / Interpolation}
\label{sec:jdi}
Having developed Theorem\;\ref{thm:denoiser} and Theorem\;\ref{thm:inter} to relate linear denoiser $\bPsi$ and linear interpolator $\bTheta$ to graph-prior MAP problems \eqref{eq:MAP_denoise} and \eqref{eq:MAP_inter} respectively, we study different joint denoising / interpolation formulations in this section.

\begin{figure}[h]
    \vspace{-1em}
    \centering
    \begin{minipage}{0.48\linewidth}
        \centering
        \includegraphics[width=\linewidth]{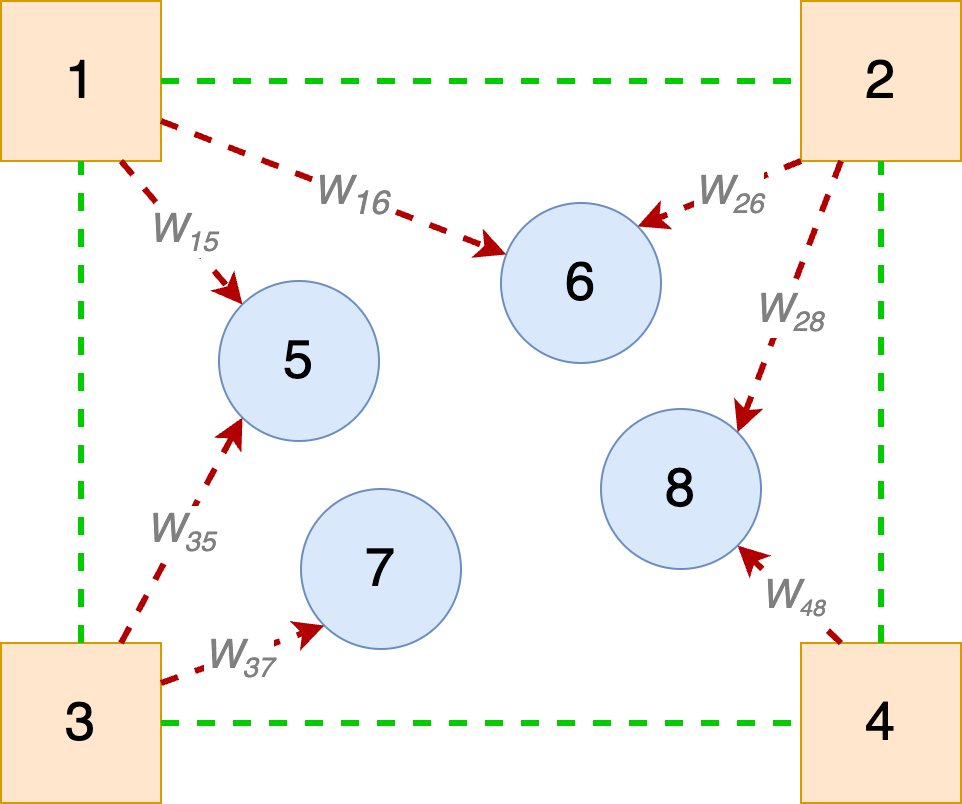}
        \subcaption{}
    \end{minipage}
    \hfill
    \begin{minipage}{0.48\linewidth}
        \centering
        \includegraphics[width=\linewidth]{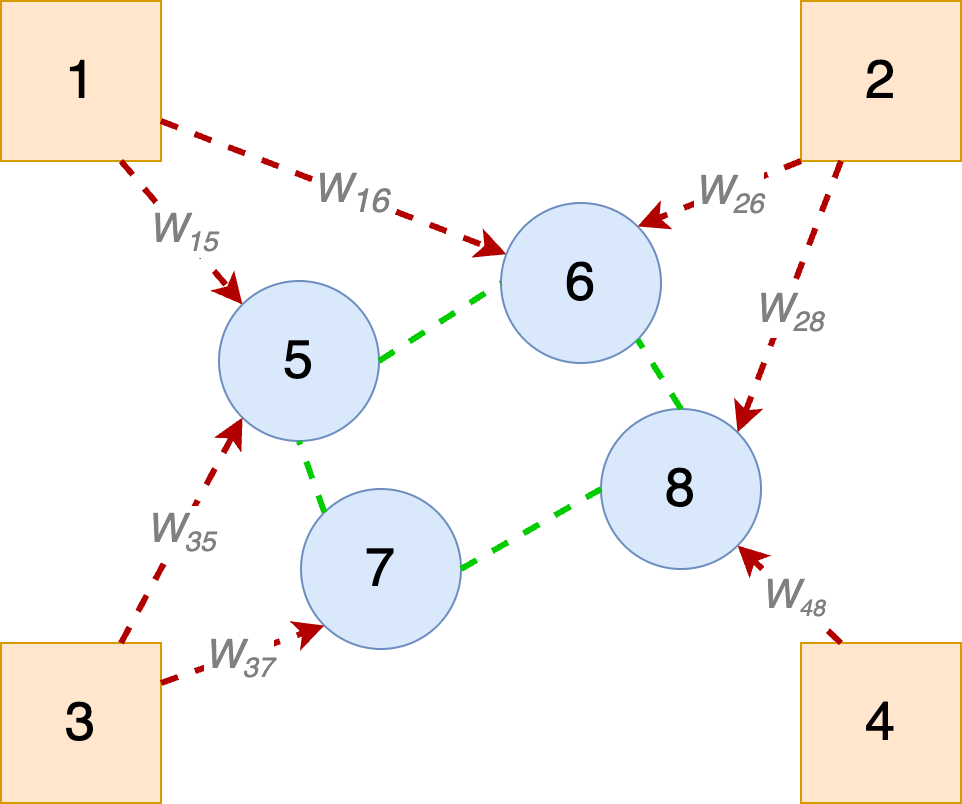}
        \subcaption{}
    \end{minipage}
    \vspace{-0.1in}
    \captionsetup{font=scriptsize}
    \caption{Illustrations of mixed graphs for joint interpolation / denoising. Pixels 5,6,7,8 are interpolated using pixels 1,2,3,4. Directed edges for interpolation are shown as red dashed arrows, while undirected edges for denoising are shown as green dashed lines. (a) corresponds to corollary\;\ref{corol:separable} where input pixels are denoised before interpolation. (b) corresponds to corollary\;\ref{corol:non-separable} where interpolated pixels are denoised after interpolation.}
    \label{fig:image2}
    \vspace{-2em}
\end{figure}

\subsection{Joint Formulation with Separable Solution}

Denote by $\A \in \mathbb{R}^{(M+N) \times (M+N)}$ in the form \eqref{eq:A_directed} an adjacency matrix of a \textit{directed} graph $\cG^d$ connecting $M$ original pixels to $N$ new pixels, corresponding to an interpolator $\bTheta$. 
Further, denote by $\L \in \mathbb{R}^{M \times M}$ a graph Laplacian matrix for an \textit{undirected} graph $\cG^u$ inter-connecting the $M$ noisy original pixels, corresponding to a denoiser $\bPsi$. 
One direct formulation for joint denoising / interpolation is to simply combine terms in MAP objectives \eqref{eq:MAP_denoise} and \eqref{eq:MAP_inter} as

\vspace{-0.05in}
\begin{small}
\begin{align}
\min_{\x} \|\y - \H \x\|^2_2 + \gamma \| \H(\x-\A\x)\|^2_2 + \mu (\H \x)^{\top} \L (\H \x) .
\label{eq:MAP_mixed}
\end{align}
\end{small}\noindent
In words, \eqref{eq:MAP_mixed} states that sought signal $\x$ should be smooth w.r.t. \textit{two} graphs $\cG^d$ and $\cG^u$: i) $\x$ should be similar to its shifted version $\A \x$ where $\A$ is an adjacency matrix for a \textit{directed} graph $\cG^d$, and ii) original pixels $\H \x$ should be smooth w.r.t. to an \textit{undirected} graph $\cG^u$ defined by $\L$. 
We show that the optimal solution to the posed MAP problem \eqref{eq:MAP_mixed} takes a particular form.
\begin{corollary}
\eqref{eq:MAP_mixed} has a separable solution. 
\label{corol:separable}
\end{corollary}

\begin{proof}
Optimization \eqref{eq:MAP_mixed} is an unconstrained convex QP problem.
Taking the derivative w.r.t. $\x$ and setting it to $0$, we get

\vspace{-0.05in}
\begin{footnotesize}
\begin{align}
\left( \H^{\top} \H + \gamma \left( (\I-\A)^\top \H^{\top} \H (\I - \A) \right) + \mu \H^{\top} \L \H \right) \x^* = \H^{\top} \y .
\label{eq:MAP_mixed_linSys}
\end{align}
\end{footnotesize}\noindent
Denote by $\C$ the coefficient matrix on the left-hand side.
Given $\H = [\I_M ~ \0_{M,N}]$, $\H^\top \L \H$ has nonzero term $\L$ only in the upper-left sub-matrix. 
Given \eqref{eq:MAP_inter_linSys2}, $\C$ differs only in the upper-left block, \ie, 
\begin{align}
\C = \left[ \begin{array}{cc}
(\gamma+1) \I_M + \mu \L & -\gamma \A_{M,N} \\
-\gamma \A_{M,N}^\top & \gamma \A^2_{N,N}
\end{array} \right] .
\label{eq:coeff2_mixed}
\end{align}

Using the matrix inverse formula \eqref{eq:inverse}, we can first write block $\dot{\P} = (\dot{\A} - \dot{\B} \dot{\D}^{-1} \dot{\C})^{-1}$ as:

\vspace{-0.1in}
\begin{small}
\begin{align}
\dot{\P} &= \left( (\gamma+1)\I_M + \mu \L - (-\gamma \A_{M,N}) (\gamma \A_{N,N}^2)^{-1} (-\gamma \A_{M,N}^\top) \right)^{-1}  
\nonumber \\
&= \left( (\gamma+1)\I_M + \mu \L - \gamma \I_M \right)^{-1} = (\I_M + \mu \L)^{-1} .
\end{align}
\end{small}

Thus, the solution $\x^*$ for optimization \eqref{eq:MAP_mixed_linSys} is
\begin{align}
\x^* &= \C^{-1} \H^\top \y 
\nonumber \\
&= 
\left[ \begin{array}{c}
(\I_M + \mu \L)^{-1} \\
- (\gamma \A_{N,N}^2)^{-1} (-\gamma \A_{M,N}^\top) (\I_M + \mu \L)^{-1}
\end{array} \right]
\nonumber \\
&= \left[ \begin{array}{c}
(\I_M + \mu \L)^{-1} \\
\A_{M,N}^{-1} (\I_M + \mu \L)^{-1}
\end{array} \right] = 
\left[ \begin{array}{c}
\bPsi \\
\bTheta \, \bPsi
\end{array} \right] \y,
\label{eq:MAP_mixed_linSys2}
\end{align}
where $\bPsi = \left(\I_M + \mu \L \right)^{-1}$ is the denoiser, and $\bTheta = \A_{M,N}^{-1}$ is the interpolator. 
We see that solution \eqref{eq:MAP_mixed_linSys2} to optimization \eqref{eq:MAP_mixed} is \textit{entirely separable}: input $\y$ first undergoes denoising via original denoiser $\bPsi$, then subsequently interpolation via original interpolator $\bTheta$.
\end{proof}

\subsection{Joint Formulation with Non-separable Solution} 

Next, we consider a scenario where we introduce a GLR smoothness prior \cite{pang17} for the $N$ interpolated pixels instead of a smoothness prior for the $M$ original pixels in \eqref{eq:MAP_mixed}, resulting in
\begin{align}
\min_{\x} & \|\y - \H \x\|^2_2 + \gamma \| \H (\x - \A \x)\|^2_2 + \kappa (\G \x)^\top \bar{\L} (\G \x),
\label{eq:MAP_mixed2}
\end{align}
where $\G = [\0_{N,M} \; \I_{N}]$ selects only the $N$ new pixels from signal $\x$, and $\bar{\L} \in \mathbb{R}^{N \times N}$ denotes a graph Laplacian matrix for an undirected graph $\bar{\cG}^u$ connecting the interpolated pixels. 

\begin{corollary}
\eqref{eq:MAP_mixed2} has a non-separable solution.
\label{corol:non-separable}
\end{corollary}

\begin{proof}
\eqref{eq:MAP_mixed2} is convex, quadratic and differentiable. 
The optimal solution $\x^*$ can be computed via a system of linear equations,

\vspace{-0.05in}
\begin{small}
\begin{align}
\left( \H^{\top} \H + \gamma \left( (\I - \A)^\top \H^\top \H (\I - \A) \right) + \cL \right) \x^* = \H^{\top} \y ,
\label{eq:MAP_mixed2_linSys}
\end{align}
\end{small}\noindent
where $\cL = \kappa \G^\top \bar{\L} \G$ and is block-diagonal, \ie,
\begin{align}
\cL = \left[ \begin{array}{cc}
\0_M & \0_{M,N} \\
\0_{N,M} & \kappa \bar{\L}
\end{array} \right] .
\end{align}
A similar derivation shows that coefficient matrix $\C$ changes from \eqref{eq:coeff2_mixed} to
\vspace{-0.1in}
\begin{align}
\C = \left[ \begin{array}{cc}
(\gamma+1) \I_M & -\gamma \A_{M,N} \\
-\gamma \A_{M,N}^\top & \gamma \A^2_{N,N} + \kappa \bar{\L}
\end{array} \right] .
\label{eq:coeff2_mixed_mod}
\end{align}

Using again the matrix inverse formula \eqref{eq:inverse}, we first write block $\dot{\P} = (\dot{\A} - \dot{\B} \dot{\D}^{-1} \dot{\C})^{-1}$ as

\vspace{-0.1in}
\begin{footnotesize}
\begin{align}
\dot{\P} &= \left( (\gamma+1) \I_M - (-\gamma \A_{M,N}) (\gamma \A_{N,N}^2 + \kappa \bar{\L} )^{-1} (-\gamma \A_{M,N}^\top)
\right)^{-1} 
\nonumber \\
&= \left( (\gamma+1) \I_M - \gamma^2 \A_{M,N} (\gamma \A_{N,N}^2 + \kappa \bar{\L} )^{-1} \A_{M,N}^\top
\right)^{-1} .
\end{align}
\end{footnotesize}\noindent 
The solution for $\x^*$ in \eqref{eq:MAP_mixed2_linSys} is

\vspace{-0.1in}
\begin{small}
\begin{align}
\x^* &= \C^{-1} \H^\top \y 
= \left[ \begin{array}{c}
\dot{\P} \\
-(\gamma \A_{N,N}^2 + \kappa \bar{\L})^{-1} (-\gamma \A_{M,N}^\top) \dot{\P}
\end{array} \right] 
\nonumber \\
&= \left[ \begin{array}{c}
\dot{\P} \\
\gamma (\gamma \A_{N,N}^2 + \kappa \bar{\L})^{-1} \A_{M,N}^\top \dot{\P}
\end{array} \right] =
\left[ \begin{array}{c}
\bPsi^* \\
\bTheta^* \, \bPsi^*
\end{array} \right] \y,
\label{eq:MAP_mixed2_linSys2}
\end{align}
\end{small}\noindent
where $\bPsi^* = \dot{\P}$ is a derived denoiser for original pixels, and $\bTheta^* = \gamma (\gamma \A_{N,N}^2 + \kappa \bar{\L})^{-1} \A_{M,N}^\top$ is an augmented interpolator.
Since $\bTheta^* \bPsi^* \neq \bar{\bPsi} \bTheta$, the solution to \eqref{eq:MAP_mixed2} is not separable\footnote{Though the solution $\bTheta^* \bPsi^*$ to \eqref{eq:MAP_mixed2} is a sequence of two separate matrix operations, each matrix is a non-separable function of input matrices $\{\A, \bar{\L}\}$ or $\{\bTheta, \bar{\bPsi}\}$.}.
\end{proof}

\vspace{0.05in}
\noindent
\textbf{Remarks}: 
\eqref{eq:MAP_mixed2_linSys2} shows not only that the solution to \eqref{eq:MAP_mixed2} is non-separable, but exactly how the optimal filters $\bTheta^* \bPsi^*$ are computed.
$\bPsi^*$ is a denoiser because it denoises $M$ original pixels.
$\bTheta^*$ is an interpolator because it operates on the $M$ denoised original pixels and outputs $N$ interpolated pixels. 
$\bPsi^*$ and $\bTheta^*$ are derivative denoising and interpolation operators that are computable functions of original directed graph adjacency matrix $\A$ for interpolation and undirected graph Laplacian $\bar{\L}$ for denoising. 
We show next that using the derived operators for joint interpolation / denoising can result in performance better than separate schemes.



\vspace{-0.05in}
\section{Experiments}
\label{sec:results}
\vspace{-0.05in}
\subsection{Experimental setup}

\vspace{-0.05in}
Experiments were conducted to test our derived operators in \eqref{eq:MAP_mixed2_linSys2} for joint denoising / interpolation (denoted by \texttt{joint}) compared to original sequential operations (denoted by \texttt{sequential}) in the non-separable case. 
For denoisers, we employed Gaussian filter \cite{milanfar13}, bilateral filter (BF) \cite{tomasi98} and nonlocal means (NLM) \cite{buades2011non, buades05}.
For interpolators, we used linear operators for image rotation and warping using Homography transform. 
The output images were evaluated using signal-to-noise ratio (PSNR). 
Popular $512 \times 512$ grayscale images, \texttt{Lena} and \texttt{peppers}, were used for rotation and warping, respectively. The experiments were run in Matlab R2022a\footnote{The code developed for these experiments are made available at \href{https://github.com/sybernix/icassp24-joint}{our GitHub repository}}. 
Gaussian noise of different variances were added to the images.

The joint denoising / interpolation operation was performed on $10 \times 10$ output patches. 
The size and location of the input patch from which the output patch is generated depend on the interpolation operation and the output location. 
To ensure that the number of input pixels is equal to the output pixels (\ie, $M = N$), we interpolated dummy pixels by adding rows to $\bTheta$. 
It is also important to ensure that $\bTheta$ is full rank (thus invertible) when adding new rows.

To ensure denoiser $\bPsi$ is non-expansive, symmetric and PD according to Theorem\;\ref{thm:denoiser}, we ran the Sinkhorn-Knopp procedure \cite{knight2008sinkhorn} for an input linear denoiser with non-negative entries and independent rows, so that matrix $\bPsi$ was \textit{double stochastic}, and thus its eigenvalues satisfied $|\lambda_i| \leq 1, \forall i$. 
Note that for a chosen interpolation operation, $\bTheta$ changed from patch to patch, and so the denoiser needed to adapt to the input and output dimensions when using the Gaussian filter. 
For BF, and NLM, the denoiser itself changed from patch to patch. 
To solve \eqref{eq:MAP_mixed2_linSys} in each iteration, \texttt{pcg} function in MATLAB implementing a version of conjugate gradient \cite{CG} was used.

Image rotation was performed at 20 degrees anti-clockwise, and for image warping a homography matrix of [1, 0.2, 0; 0.1, 1, 0; 0, 0, 1] was used. For the experiments with Gaussian filter and BF, the hyperparameters were selected as $\mu = 0.3$, $\gamma = 0.5$, $\kappa = 0.3$. For Gaussian denoiser, a variance of $0.3$ was used, and for BF, variance of $0.3$ was used for both spatial and range kernels. For experiments with NLM, $\gamma = 0.6$, $\kappa = 0.2$, while $\mu$ was kept the same. 
The patch size for NLM was $3 \times 3$ and the search window size was $9 \times 9$.

\vspace{-0.05in}
\subsection{Experimental Results}

\begin{figure}
    \centering
    \begin{minipage}{0.49\columnwidth}
        \centering
        \begin{tikzpicture}
        \begin{axis}[
            title={(a)},
            title style={yshift=-1.5ex,font=\bfseries},
            xlabel={Noise Variance (x$10^{-2}$)},
            ylabel={PSNR (dB)},
            width=1.2\linewidth,
            height=3.5cm,
            enlarge x limits=false,
            grid=major,
            legend pos=north east,
            xticklabel style={font=\tiny},
            yticklabel style={font=\tiny},
            ylabel near ticks,
            legend style={font=\fontsize{5pt}{5pt}\selectfont,draw=none},
            xlabel style={
                font=\fontsize{7pt}{5pt}\selectfont, 
                at={(0.5, 0.2)}, 
                anchor=north
            },
            ylabel style={
                font=\fontsize{6pt}{5pt}\selectfont, 
                at={(-0.15, 0.7)}, 
                anchor=east
            },
            font=\small
        ]
        \addplot[blue, mark=*, mark size=1pt] coordinates {
            (2, 21.587485)
            (2.5, 20.90432783)
            (3, 20.31921832)
            (3.5, 19.83167639)
            (4, 19.42715716)
            (4.5, 19.022326)
            (5, 18.71150304)
            (5.5, 18.41532741)
            (6, 18.14750901)
            (6.5, 17.89598423)
            (7, 17.66740648)
            (7.5, 17.43878226)
            (8, 17.23422867)
            (8.5, 17.07907882)
            (9, 16.91256899)
            (9.5, 16.75608323)
            (10, 16.60490031)
        };
        \addlegendentry{Joint} 
    
        \addplot[red, mark=square*, mark size=1pt] coordinates {
            (2, 20.73798887)
            (2.5, 19.92855724)
            (3, 19.27870153)
            (3.5, 18.72884451)
            (4, 18.26982873)
            (4.5, 17.82756754)
            (5, 17.47745643)
            (5.5, 17.15760775)
            (6, 16.86586809)
            (6.5, 16.60124918)
            (7, 16.35937255)
            (7.5, 16.13717938)
            (8, 15.91209673)
            (8.5, 15.74208065)
            (9, 15.56832878)
            (9.5, 15.40428717)
            (10, 15.25117123)
        };
        \addlegendentry{Sequential} 
        \end{axis}
        \end{tikzpicture}
    \end{minipage}
    \hfill
    \begin{minipage}{0.49\columnwidth}
        \centering
        \begin{tikzpicture}
        \begin{axis}[
            title={(b)},
            title style={yshift=-1.5ex,font=\bfseries},
            xlabel={Noise Variance (x$10^{-2}$)},
            ylabel={PSNR (dB)},
            width=1.2\linewidth,
            height=3.5cm,
            enlarge x limits=false,
            grid=major,
            legend pos=north east,
            xticklabel style={font=\tiny},
            yticklabel style={font=\tiny},
            ylabel near ticks,
            legend style={font=\fontsize{5pt}{5pt}\selectfont,draw=none},
            xlabel style={
                font=\fontsize{7pt}{5pt}\selectfont, 
                at={(0.5, 0.2)}, 
                anchor=north
            },
            ylabel style={
                font=\fontsize{6pt}{5pt}\selectfont, 
                at={(-0.15, 0.7)}, 
                anchor=east
            },
            font=\small
        ]
        \addplot[blue, mark=*, mark size=1pt] coordinates {
            (12.5, 20.53081724)
            (15, 19.95725104)
            (17.5, 19.45396617)
            (20, 19.01517752)
            (22.5, 18.64232406)
            (25, 18.2745969)
            (27.5, 18.02970606)
            (30, 17.7809938)
            (32.5, 17.55870989)
            (35, 17.31810037)
            (37.5, 17.1376074)
            (40, 17.01345504)
            (42.5, 16.81894122)
            (45, 16.73037607)
        };
        \addlegendentry{Joint} 
    
        \addplot[red, mark=square*, mark size=1pt] coordinates {
            (12.5, 20.24606572)
            (15, 19.53212844)
            (17.5, 18.98609857)
            (20, 18.47522446)
            (22.5, 18.04511827)
            (25, 17.66611801)
            (27.5, 17.37657613)
            (30, 17.10309109)
            (32.5, 16.86005929)
            (35, 16.59374565)
            (37.5, 16.41857797)
            (40, 16.2698268)
            (42.5, 16.05460114)
            (45, 15.95740103)
        };
        \addlegendentry{Sequential} 
        \end{axis}
        \end{tikzpicture}
    \end{minipage}

    \begin{minipage}{0.49\columnwidth}
        \centering
        \begin{tikzpicture}
        \begin{axis}[
            title={(c)},
            title style={yshift=-1.5ex,font=\bfseries},
            xlabel={Noise Variance (x$10^{-2}$)},
            ylabel={PSNR (dB)},
            width=1.2\linewidth,
            height=3.5cm,
            enlarge x limits=false,
            grid=major,
            legend pos=north east,
            xticklabel style={font=\tiny},
            yticklabel style={font=\tiny},
            ylabel near ticks,
            font=\small,
            legend style={font=\fontsize{5pt}{5pt}\selectfont,draw=none},
            xlabel style={
                font=\fontsize{7pt}{5pt}\selectfont, 
                at={(0.5, 0.2)}, 
                anchor=north
            },
            ylabel style={
                font=\fontsize{6pt}{5pt}\selectfont, 
                at={(-0.15, 0.7)}, 
                anchor=east
            }
        ]
        \addplot[blue, mark=*, mark size=1pt] coordinates {
            (2, 22.85433434)
            (2.5, 21.9958582)
            (3, 21.32414268)
            (3.5, 20.79961061)
            (4, 20.31556308)
            (4.5, 19.87632938)
            (5, 19.50303237)
            (5.5, 19.16995014)
            (6, 18.87852137)
            (6.5, 18.57946564)
            (7, 18.36629438)
            (7.5, 18.13947484)
            (8, 17.90983536)
            (8.5, 17.71168274)
            (9, 17.51181452)
            (9.5, 17.3721067)
            (10, 17.19671591)
        };
        \addlegendentry{Joint} 
    
        \addplot[red, mark=square*, mark size=1pt] coordinates {
            (2, 22.11974936)
            (2.5, 21.1869656)
            (3, 20.45961032)
            (3.5, 19.89594629)
            (4, 19.38216808)
            (4.5, 18.92146386)
            (5, 18.53060917)
            (5.5, 18.17274666)
            (6, 17.87582365)
            (6.5, 17.57251431)
            (7, 17.35475666)
            (7.5, 17.11837766)
            (8, 16.87093696)
            (8.5, 16.68271421)
            (9, 16.47874404)
            (9.5, 16.33576235)
            (10, 16.14892135)
        };
        \addlegendentry{Sequential} 
        \end{axis}
        \end{tikzpicture}
    \end{minipage}
    \hfill
    \begin{minipage}{0.49\columnwidth}
        \centering
        \begin{tikzpicture}
        \begin{axis}[
            title={(d)},
            title style={yshift=-1.5ex,font=\bfseries},
            xlabel={Noise Variance (x$10^{-2}$)},
            ylabel={PSNR (dB)},
            width=1.2\linewidth,
            height=3.5cm,
            enlarge x limits=false,
            grid=major,
            legend pos=north east,
            xticklabel style={font=\tiny},
            yticklabel style={font=\tiny},
            ylabel near ticks,
            legend style={font=\fontsize{5pt}{5pt}\selectfont,draw=none},
            font=\small,
            xlabel style={
                font=\fontsize{7pt}{5pt}\selectfont, 
                at={(0.5, 0.2)}, 
                anchor=north
            },
            ylabel style={
                font=\fontsize{6pt}{5pt}\selectfont, 
                at={(-0.15, 0.7)}, 
                anchor=east
            }
        ]
        \addplot[blue, mark=*, mark size=1pt] coordinates {
            (2, 22.88869845)
            (2.5, 22.09556211)
            (3, 21.44567393)
            (3.5, 20.85862609)
            (4, 20.39798979)
            (4.5, 19.99750479)
            (5, 19.6214151)
            (5.5, 19.27550648)
            (6, 19.02261291)
            (6.5, 18.70546002)
            (7, 18.48461185)
            (7.5, 18.23218625)
            (8, 18.04305553)
            (8.5, 17.81840038)
            (9, 17.65993033)
            (9.5, 17.47620249)
            (10, 17.35934522)
        };
        \addlegendentry{Joint} 
    
        \addplot[red, mark=square*, mark size=1pt] coordinates {
            (2, 22.12244712)
            (2.5, 21.22696602)
            (3, 20.49146135)
            (3.5, 19.87125975)
            (4, 19.38107537)
            (4.5, 18.97335459)
            (5, 18.56215273)
            (5.5, 18.19682779)
            (6, 17.94406525)
            (6.5, 17.59528876)
            (7, 17.38390242)
            (7.5, 17.11846249)
            (8, 16.91795141)
            (8.5, 16.70790565)
            (9, 16.52024553)
            (9.5, 16.34767813)
            (10, 16.22246286)
        };
        \addlegendentry{Sequential} 
        
        \end{axis}
        \end{tikzpicture}
    \end{minipage}
    \captionsetup{font=scriptsize}
    \vspace{-1em}
    \caption{Plots comparing the performance of  \texttt{joint} vs \texttt{sequential} over a range of noise variance using 4 combinations of interpolators and denoisers. \textbf{(a)} Image rotation + Bilateral denoiser. \textbf{(b)} Image rotation + Non-Local Means denoiser. \textbf{(c)} Image warping + Bilateral denoiser. \textbf{(d)} Image warping + Gaussian denoiser.}
    \vspace{-2em}
    \label{fig:side_by_side_plots}
\end{figure}
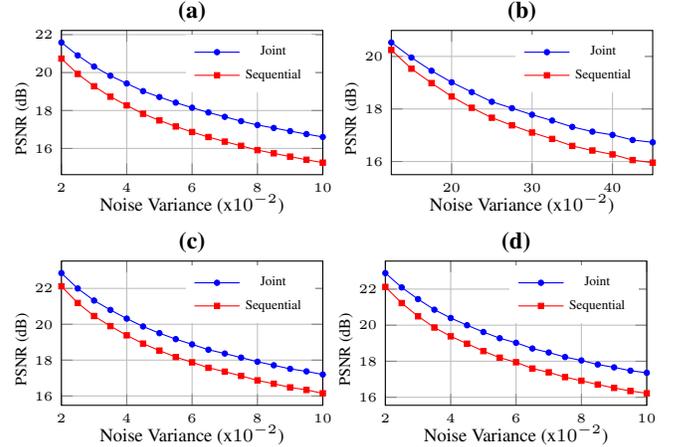

\vspace{-0.05in}
Fig.\;\ref{fig:side_by_side_plots} show that \texttt{joint} performed better than \texttt{sequential} in general. 
While the performance of both schemes degraded as noise variance increased, the performance of \texttt{sequential} degraded faster than \texttt{joint}. 
In the experiment with image rotation and Bilateral denoiser, we observe a maximum PSNR gain of $1.35$dB, and when NLM was used, the maximum gain was $0.77$dB. Note that we have reported results for NLM over a larger range of noise variance, because NLM generally produced high-quality output, and thus the PSNR difference between \texttt{joint} and \texttt{sequential} is small  at low noise levels. 
For image warping, the maximum gains were $1.05$dB and $1.14$dB for bilateral and Gaussian denoisers, respectively.

\vspace{-0.05in}
\section{Conclusion}
\label{sec:conclude}
\vspace{-0.05in}
We presented two theorems, under mild conditions, connecting any linear denoiser / interpolator to optimal graph filter using undirected / directed graph smoothness priors, respectively. 
The theorems demonstrate the generality of graph filters and provide graph interpretations for common linear denoisers / interpolators. 
Using the two theorems, we examine scenarios of joint denoising / interpolation where the optimal solution can be separable or non-separable. 
In the latter case, we analytically derive derivative denoiser / interpolator computed as functions of original denoiser and interpolator. 
We demonstrate that using these computed operators resulted in noticeable performance gain over seperate schemes in a range of joint denoising / interpolation settings.

\bibliographystyle{IEEEbib}
\bibliography{strings,refs}

\end{document}